\documentclass[aps,onecolumn,letterpaper,oneside,preprint,tightenlines,draft,%
nobibnotes,nofootinbib,amsfonts,amssymb,amsmath,showpacs,showkeys]{revtex4}
\newcommand{\fF}{\mathfrak{F}}
\newcommand{\cN}{\mathcal{N}}
\newcommand{\cP}{\mathcal{P}}
\newcommand{\bH}{\boldsymbol{H}}
\newcommand{\bQ}{\boldsymbol{Q}}
\newcommand{\rmd}{\mathrm{d}}%neIOPART
\newcommand{\del}{\partial}

\begin{document}

%\markboth{Authors' Names}
%{Instructions for Typing Manuscripts (Paper's Title)}%WSPC

%\catchline{}{}{}{}{}%WSPC
%
%%%%%%%%%%%%%%%%%%%%%%%%%%%%%%%%%%%%%%%%%%%%%%%%%%%%%%%%%%%%%%%%%%%%

\title{$\cN$-fold Supersymmetric Quantum Mechanics with Reflections}
%\article[Short title]{TYPE}{Full title}%IOPART
%ELSART,WSPC
%\author{},
%\ead{}%ELSART
%\address{}
%\author{Toshiaki Tanaka}
%\ead{toshiaki@post.kek.jp}%ELSART
%\address{Institute of Particle and Nuclear Studies,
% High Energy Accelerator Research Organization (KEK),
% 1-1 Oho, Tsukuba, Ibaraki 305-0801, Japan}
%REVTEX4
\author{Toshiaki Tanaka}
\email{toshiaki@post.kek.jp}
\affiliation{Institute of Particle and Nuclear Studies,
 High Energy Accelerator Research Organization (KEK),
 1-1 Oho, Tsukuba, Ibaraki 305-0801, Japan}
%  \altaffiliation{}
%IOPART
%\author{Author 1\dag, Author 2\ddag and Toshiaki Tanaka\S}
%\address{\dag Address 1}
%\address{\ddag Address 2}
%\address{\S Institute of Particle and Nuclear Studies,
% High Energy Accelerator Research Organization (KEK),
% 1-1 Oho, Tsukuba, Ibaraki 305-0801, Japan}
%\ead{toshiaki@post.kek.jp}
%\eads{\mailto{email 1}, \mailto{email 2},
% \mailto{toshiaki@post.kek.jp}}

%\date{\today}

\begin{abstract}

We formulate $\cN$-fold supersymmetry in quantum mechanical systems
with reflection operators. As in the cases of other systems, they possess the two
significant characters of $\cN$-fold supersymmetry, namely, almost isospectrality
and weak quasi-solvability. We construct explicitly the most general $1$- and
$2$-fold supersymmetric quantum mechanical systems with reflections. In the case
of $\cN=2$, we find that there are seven inequivalent such systems, three of which
are characterized by three arbitrary functions having definite parity while
the other four of which are by two.
In addition, four of the seven inequivalent systems do not reduce to ordinary
quantum systems without reflections. Furthermore, in certain particular cases,
they are essentially equivalent to the most general two-by-two Hermitian matrix
$2$-fold supersymmetric quantum systems obtained previously by us.

%\keywords{keyword1; keyword2; keyword3.}%WSPC
\end{abstract}

%\ccode{PACS Nos.: nn.nn.Xx; nn.nn.XX}%WSPC

%REVTEX4
%REVTEX4

%\begin{keyword}%ELSART
%  keyword 1\sep keyword 2\sep keyword 3
% \sep keyword 4\sep keyword 5\sep keyword 6
% \PACS nn.nn.Xx\sep nn.nn.Xx\sep nn.nn.Xn\sep nn.nn.Xx
%\end{keyword}%ELSART

%\pacs{nn.nn.Xx, nn.nn.Xx, nn.nn.Xx, nn.nn.Xx}%IOPART
%\submitto{Journal}

\preprint{TH-1511}%REVTEX4

\maketitle

\section{Introduction}
\label{sec:intro}

Recently, supersymmetry (SUSY) was formulated for one-dimensional quantum
mechanical systems with reflections in Ref.~\cite{PVZ11}. One of its characteristic
features is that both a supersymmetric Hamiltonian and a supercharge component
involve reflection operators. An intriguing aspect shown in this work is that exact
eigenfunctions of such a system are expressed in terms of little $-1$ Jacobi
polynomials which is one of a ``missing'' family of classical orthogonal
polynomials~\cite{VZ11c}. Hence, it is interesting to study what kind of
Hamiltonians involving reflection operators admit exact eigenfunctions
which are expressible in terms of such a ``missing'' classical orthogonal
polynomial system.

On the other hand, the framework of $\cN$-fold SUSY~\cite{AIS93,AST01b,AS03}
has been shown to be quite fruitful among several generalizations of ordinary SUSY
especially since the establishment of its equivalence with weak quasi-solvability
in Ref.~\cite{AST01b}, for a review see Ref.~\cite{Ta09}. Until now, four
different types have been established, namely, type A~\cite{AST01a,Ta03a},
type B~\cite{GT04}, type C~\cite{GT05}, and type $X_{2}$~\cite{Ta10a}.
We note that almost all the models having essentially the same symmetry as
$\cN$-fold SUSY but called with other terminologies in the literature, such as
P\"{o}schl--Teller and Lam\'{e} potentials, are actually particular cases of type
A $\cN$-fold SUSY. To avoid confusion, we also note that $\cN$-fold SUSY is
different from \emph{nonlinear SUSY} which has been long employed since the work
by Samuel and Wess~\cite{SW83} in 1983 to indicate nonlinearly realized SUSY
originated from the work by Akulov and Volkov~\cite{VA72} in 1972. For recent
works on nonlinear SUSY, see, e.g., Ref.~\cite{ADGT10} and references cited therein.
Due to the facts that the $\cN=1$ case corresponds to ordinary SUSY and that exact
solvability always means weak quasi-solvability, the framework of $\cN$-fold SUSY
enables us to formulate systematically ordinary SUSY and exactly solvable quantum
systems as its particular cases. In fact, we successfully formulated in Ref.~\cite{Ta12a}
$\cN$-fold SUSY in quantum mechanical matrix models as a generalization of ordinary
SUSY quantum mechanical matrix models in, e.g., Ref.~\cite{NK11} and references cited
therein. Hence, it is quite natural to ask whether a formulation of $\cN$-fold SUSY
is possible for quantum mechanical systems with reflection operators. To the best
of our knowledge, there have been no such attempts in the existing literature.

In this article, we formulate for the first time $\cN$-fold SUSY for quantum
mechanical systems with reflection operators for all positive integral $\cN$.
To see concretely what kinds of forms such systems must have,
we construct the most general $1$- and $2$-fold SUSY systems by solving
directly all the conditions for the respective SUSYs. In the case of $\cN=1$, we
find in particular that ordinary SUSY algebra can be realized under a less restrictive
condition than the one presupposed in Ref.~\cite{PVZ11}. In the case of $\cN=2$,
we find that there are seven inequivalent systems, three of which are characterized
by three arbitrary functions having definite parity while the other four of which are
by two. In addition, we also find that four of the seven inequivalent systems do not
admit a reduction to $2$-fold SUSY ordinary quantum systems without reflection
operators. Furthermore, in certain particular cases, they are essentially equivalent
to the most general two-by-two Hermitian matrix $2$-fold supersymmetric quantum
systems obtained previously by us in Ref.~\cite{Ta12a}.

We organize this article as follows. In the next section, we first summarize
fundamental formulas which are frequently needed for calculations involving both
differential and reflection operators. Then, we generically define $\cN$-fold SUSY
in quantum mechanical systems with reflections. In Section~\ref{sec:1f}, we
present the most general results in the $\cN=1$ case which corresponds to
ordinary SUSY. We also clarify the relation between our formalism and the SUSY QM
with reflections formulated in Ref.~\cite{PVZ11}. In Section~\ref{sec:2f}, we
investigate in detail the $\cN=2$ case. We explicitly solve all the conditions for
$2$-fold SUSY to obtain the most general form of the latter systems.
In the last section, we refer to several future issues to be
followed after this work.

\section{Preliminaries and General Setting}
%\label{sec:gen}
%\nosections%IOPART
First of all, let $\cP$ denote a reflection or parity operator whose action on
an element of a linear function space $\fF$ is defined by
\begin{align}
\cP\cdot\psi (q)=\psi (-q):=\psi_{\cP}(q).
\label{eq:defP}
\end{align}
We will hereafter use the last notation $\psi_{\cP}$ frequently especially when
we will omit the argument of the function under consideration.
On the other hand, if $f(q)$ is a multiplicative operator in $\fF$, we have instead
an operator relation as
\begin{align}
\cP f(q)=f_{\cP}(q)\cP.
\end{align}
Another important operator relation is the one between reflection and differential
operators
\begin{align}
\frac{\rmd}{\rmd q}\cP=-\cP\frac{\rmd}{\rmd q}.
\label{eq:dP1}
\end{align}
Due to the latter anti-commutativity, we have in particular
\begin{align}
(\psi_{\cP})'(q)=\frac{\rmd\psi_{\cP}(q)}{\rmd q}
 =-\cP\cdot\frac{\rmd\psi(q)}{\rmd q}=-(\psi')_{\cP}(q).
\label{eq:dP2}
\end{align}
Any function $f(q)$ admits the decomposition into its even and odd parts, denoted
respectively by $f_{+}(q)$ and $f_{-}(q)$, as
\begin{align}
f(q)=f_{+}(q)+f_{-}(q),\qquad 2f_{\pm}(q)=f(q)\pm f_{\cP}(q).
\label{eq:evod}
\end{align}
It is evident from (\ref{eq:dP2}) and (\ref{eq:evod}) that
\begin{align}
(f')_{\pm}(q)=(f_{\mp})'(q).
\label{eq:dpm}
\end{align}
The formulas (\ref{eq:defP})--(\ref{eq:dpm}) are fundamental tools for dealing
with differential operators with reflections.

A quantum mechanical system we shall consider here is a pair of Schr\"{o}dinger
operators which involve reflection operators as follows:
\begin{align}
H^{\pm}=-\frac{1}{2}\frac{\rmd^{2}}{\rmd q^{2}}+V_{0}^{\pm}(q)+V_{1}^{\pm}(q)\cP,
\label{eq:H+-}
\end{align}
where the potential functions $V_{0}^{\pm}(q)$ and $V_{1}^{\pm}(q)$ are to be
determined later. Let us introduce a pair of linear differential operators of order
$\cN$ with reflection operators
\begin{subequations}
\label{eqs:PN+-}
\begin{align}
P_{\cN}^{-}&=\frac{\rmd^{\cN}}{\rmd q^{\cN}}+\sum_{k=0}^{\cN-1}\left[
 w_{k}^{[\cN]}(q)+v_{k}^{[\cN]}(q)\cP\right]\frac{\rmd^{k}}{\rmd q^{k}},\\
P_{\cN}^{+}&=(P_{\cN}^{-})^{\textrm{T}}=(-1)^{\cN}\frac{\rmd^{\cN}}{\rmd q^{\cN}}
 +\sum_{k=0}^{\cN-1}(-1)^{k}\frac{\rmd^{k}}{\rmd q^{k}}\left[w_{k}^{[\cN]}(q)
 +\cP v_{k}^{[\cN]}(q)\right],
\end{align}
\end{subequations}
where $w_{k}^{[\cN]}(q)$ and $v_{k}^{[\cN]}(q)$ ($k=0,\dots,\cN-1$) are in general
complex analytic functions, and the superscript T denotes transposition. We will
hereafter omit the superscript $[\cN]$ for the simplicity unless the omission may
cause confusion or ambiguity. Then, the system (\ref{eq:H+-}) is said to be
\emph{$\cN$-fold supersymmetric} with respect to (\ref{eqs:PN+-}) if the
following relations are all satisfied:
\begin{align}
&P_{\cN}^{\mp}H^{\mp}-H^{\pm}P_{\cN}^{\mp}=0,
\label{eq:NS1}\\
&P_{\cN}^{\mp}P_{\cN}^{\pm}=2^{\cN}\left[(H^{\pm}+C_{0})^{\cN}+\sum_{k=1}^{\cN-1}
 C_{k}(H^{\pm}+C_{0})^{\cN-k-1}\right],
\label{eq:NS2}
\end{align}
where $C_{k}$ ($k=0,\dots,\cN-1$) are constant multiplicative operators with reflections
\begin{align}
C_{k}=C_{k0}+C_{k1}\cP.
\label{eq:C0}
\end{align}
The two intertwining relations in (\ref{eq:NS1}) are related by transposition if both
the potential terms $V_{1}^{+}(q)$ and $V_{1}^{-}(q)$ are even and thus commute with
$\cP$, since in the latter case both the Hamiltonians $H^{+}$ and $H^{-}$ are
invariant under transposition, $(H^{\pm})^{\textrm{T}}=H^{\pm}$. As is usual,
we can express an $\cN$-fold SUSY system in a unified way by introducing
a superHamiltonian $\bH$ and a pair of $\cN$-fold supercharges
$\bQ_{\cN}^{\pm}$ as
\begin{align}
\bH=H^{-}\psi^{-}\psi^{+}+H^{+}\psi^{+}\psi^{-},\qquad
 \bQ_{\cN}^{\pm}=P_{\cN}^{\mp}\psi^{\pm},
\end{align}
where $\psi^{\pm}$ are fermionic variables satisfying $(\psi^{\pm})^{2}=0$ and
$\{\psi^{+},\psi^{-}\}=1$. Then, the $\cN$-fold SUSY relations (\ref{eq:NS1}) and
(\ref{eq:NS2}) are summarized in $\cN$-fold superalgebra
\begin{align}
\bigl[\bQ_{\cN}^{\pm}, \bH\bigr]=0,\quad\bigl\{\bQ_{\cN}^{+},\bQ_{\cN}^{-}\bigr\}=
 2^{\cN}\left[(\bH+C_{0})^{\cN}+\sum_{k=1}^{\cN-1}C_{k}(\bH+C_{0})^{\cN-k-1}\right].
\end{align}
We note that in the case of $\cN=1$ the above definition of $\cN$-fold SUSY is
slightly different from the SUSY QM with reflections in Ref.~\cite{PVZ11}.
The exact relation between our $P_{1}^{-}$ and a supercharge component $Q$
in the latter reference, Eq.~(2.11), is $P_{1}^{-}=\sqrt{2}Q\cP$ with $w_{0}=U$,
$v_{0}=V$, and $\cP=R$. In particular, our $\cN$-fold supercharge components
$P_{\cN}^{\pm}$ do not possess formal Hermiticity in contrast with $Q$ in
the latter. The relations between our $1$-fold SUSY pair of Hamiltonians and
a SUSY Hamiltonian $H$ in the latter are $H^{+}=H$ and $H^{-}=\cP H\cP$.
In particular, $H^{-}=H^{+}$ if $H^{+}$ commutes with a reflection operator $\cP$.

It is evident from the definition that $\cN$-fold SUSY quantum systems with
reflections (\ref{eq:H+-})--(\ref{eq:C0}) reduce to ones without reflections if 
$V_{1}^{\pm}(q)=v_{k}(q)=C_{01}=0$ for all $k=0,\dots,\cN-1$. As in the case without
reflections, the first relation (\ref{eq:NS1}) immediately implies almost isospectrality
of $H^{\pm}$ and \emph{weak quasi-solvability} $H^{\pm}\ker P_{\cN}^{\pm}
\subset\ker P_{\cN}^{\pm}$.

\section{Ordinary SUSY}
\label{sec:1f}

In this section, we shall examine the $\cN=1$ case, namely, ordinary SUSY QM
with reflections. Components of supercharges are given by
\begin{align}
P_{1}^{-}=\frac{\rmd}{\rmd q}+w_{0}(q)+v_{0}(q)\cP,\qquad
 P_{1}^{+}=-\frac{\rmd}{\rmd q}+w_{0}(q)+\cP v_{0}(q).
\end{align}
A direct calculation immediately yields
\begin{align}
P_{1}^{-}P_{1}^{+}&=-\frac{\rmd^{2}}{\rmd q^{2}}-2v_{0+}\cP\frac{\rmd}{\rmd q}
 +w'_{0}+(w_{0})^{2}+(v_{0})^{2}+\left(-(v'_{0})_{\cP}+w_{0}v_{0\cP}
 +w_{0\cP}v_{0}\right)\cP,\\
P_{1}^{+}P_{1}^{-}&=-\frac{\rmd^{2}}{\rmd q^{2}}+2v_{0+}\cP\frac{\rmd}{\rmd q}
 -w'_{0}+(w_{0})^{2}+(v_{0\cP})^{2}+\left(-v'_{0}+w_{0}v_{0}+w_{0\cP}v_{0\cP}\right)\cP.
\end{align}
Hence, they are of the form (\ref{eq:H+-}) if and only if
\begin{align}
2v_{0+}(q)=v_{0}(q)+v_{0\cP}(q)=0,
\end{align}
that is, $v_{0}(q)$ is an odd function $v_{0}(q)=v_{0-}(q)$. Under the latter condition,
the $\cN$-fold superalgebra (\ref{eq:NS2}) in the case of $\cN=1$ holds and the
potential terms in (\ref{eq:H+-}) are expressed as
\begin{align}
2V_{0}^{\pm}&=\pm w'_{0}+(w_{0})^{2}+(v_{0-})^{2}-2C_{00},\\
2V_{1}^{\pm}&=-(v_{0-})'\mp 2w_{0-}v_{0-}-2C_{01},
\end{align}
where $C_{00}$ and $C_{01}$ are constants defined by (\ref{eq:C0}). The intertwining
relation (\ref{eq:NS1}) is trivially satisfied.
We note that $V_{1}^{\pm}(q)$ is automatically even for an arbitrary $w_{0}(q)$.
In this respect, it is also worth mentioning that the evenness of $w_{0}(q)$ is
not inevitable for SUSY although it was presupposed in Ref.~\cite{PVZ11}. When
$w_{0}(q)$ is even,

\section{$2$-fold SUSY}
\label{sec:2f}

Next, we shall proceed to the $\cN=2$ case where components of $2$-fold
supercharges are given by
\begin{subequations}
\label{eqs:P2+-}
\begin{align}
P_{2}^{-}&=\frac{\rmd^{2}}{\rmd q^{2}}+\left[w_{1}(q)+v_{1}(q)\cP\right]
 \frac{\rmd}{\rmd q}+w_{0}(q)+v_{0}(q)\cP,\\
P_{2}^{+}&=\frac{\rmd^{2}}{\rmd q^{2}}-\frac{\rmd}{\rmd q}\left[w_{1}(q)
 +\cP v_{1}(q)\right]+w_{0}(q)+\cP v_{0}(q).
\end{align}
\end{subequations}
Before investigating the intertwining relation (\ref{eq:NS1}) for $\cN=2$,
we first note that a direct calculation shows (see Eqs.~(\ref{eq:2-2+}) and
(\ref{eq:2+2-}) in Appendix for the full formulas)
\begin{align}
P_{2}^{\mp}P_{2}^{\pm}=\del^{4}+2v_{1+}\cP\del^{3}+O(\del^{2}),
\end{align}
where $O(\del^{2})$ denotes a linear differential operator of at most second order.
Hence, it is necessary that the function $v_{1}(q)$ is odd
\begin{align}
2v_{1+}(q)=v_{1}(q)+v_{1\cP}(q)=0,
\end{align}
for satisfying the $2$-fold superalgebra (\ref{eq:NS2}), that is, $v_{1}(q)=v_{1-}(q)$.
Under the latter condition, the second-order intertwining relation $P_{2}^{-}H^{-}
-H^{+}P_{2}^{-}=0$ holds if and only if the following set of conditions are satisfied:
\begin{align}
&V_{0}^{+}-V_{0}^{-}=w'_{1},
\label{eq:co1}\\
&V_{1}^{+}-V_{1}^{-}=-(v_{1-})',
\label{eq:co2}\\
&w''_{1}+2w'_{0}+4V_{0}^{-\prime}-2w_{1}(V_{0}^{+}-V_{0}^{-})+2v_{1-}\left(
 V_{1}^{+}-(V_{1}^{-})_{\cP}\right) =0,
\label{eq:co3}\\
&(v_{1-})''-2v'_{0}-4V_{1}^{-\prime}-2v_{1-}\left( V_{0}^{+}-(V_{0}^{-})_{\cP}\right)
 -2w_{1\cP}V_{1}^{+}-2w_{1}V_{1}^{-}=0,
\label{eq:co4}\\
&w''_{0}+2V_{0}^{-\prime\prime}+2w_{1}V_{0}^{-\prime}-2v_{1-}((V_{1}^{-})_{\cP})'
 -2v_{0\cP}V_{1}^{+}+2v_{0}(V_{1}^{-})_{\cP}-2w_{0}(V_{0}^{+}-V_{0}^{-})=0,
\label{eq:co5}\\
&v''_{0}+2V_{1}^{-\prime\prime}+2w_{1}V_{1}^{-\prime}-2v_{1-}((V_{0}^{-})_{\cP})'
 -2w_{0\cP}V_{1}^{+}+2w_{0}V_{1}^{-}-2v_{0}\left( V_{0}^{+}-(V_{0}^{-})_{\cP}\right)=0.
\label{eq:co6}
\end{align}
On the other hand, using the formula
\begin{align}
4(H^{\pm})^{2}=&\;\frac{\rmd^{4}}{\rmd q^{4}}-4(V_{0}^{\pm}+V_{1}^{\pm}\cP)
 \frac{\rmd^{2}}{\rmd q^{2}}-4(V_{0}^{\pm\prime}-V_{1}^{\pm\prime}\cP)
 \frac{\rmd}{\rmd q}\notag\\
&\;-2[V_{0}^{\pm\prime\prime}-2(V_{0}^{\pm})^{2}-2V_{1}^{\pm}(V_{1}^{\pm})_{\cP}]
 -2[V_{1}^{\pm\prime\prime}-4(V_{0}^{\pm})_{+}V_{1}^{\pm}]\cP,
\end{align}
we find that the $2$-fold superalgebra $P_{2}^{\mp}P_{2}^{\pm}
=4\left[(H^{\pm}+C_{0})^{2}+C_{1}\right]$ holds for the upper sign if and only if
\begin{align}
&4V_{0}^{+}=3w'_{1}-2w_{0}+(w_{1})^{2}+(v_{1-})^{2}-4C_{00},
\label{eq:c+1}\\
&4V_{1}^{+}=-3(v_{1-})'-2v_{0+}-2w_{1-}v_{1-}-4C_{01},
\label{eq:c+2}\\
&4V_{0}^{+\prime}=3w''_{1}-2w'_{0}+2w_{1}w'_{1}+2v_{1-}(v_{1-})',
\label{eq:c+3}\\
&4V_{1}^{+\prime}=-3(v_{1-})''-2(v_{0\cP})'+2(w_{1\cP})'v_{1-}-2w_{1}(v_{1-})'
 -2w_{0-}v_{1-}-w_{1}v_{0\cP}-w_{1\cP}v_{0},
\label{eq:c+4}\\
&2V_{0}^{+\prime\prime}-4(V_{0}^{+}+C_{00})^{2}-4(V_{1}^{+}+C_{01})\left(
 (V_{1}^{+})_{\cP}+C_{01}\right) -4C_{10}=w'''_{1}-w''_{0}\notag\\
&\hspace*{30pt}+w_{1}w''_{1}+v_{1-}(v_{1-})''+w'_{1}w_{0}-w_{1}w'_{0}-(v_{1-})'v_{0}
 +v_{1-}v'_{0}-(w_{0})^{2}-(v_{0})^{2},
\label{eq:c+5}\\
&2V_{1}^{+\prime\prime}-8\left( (V_{0}^{+})_{+}+C_{00}\right)
 (V_{1}^{+}+C_{01})-4C_{11}=-(v_{1-})'''-(v_{0\cP})''+(w_{1\cP})''v_{1-}\notag\\
&\hspace*{30pt}-w_{1}(v_{1-})''-(w_{1\cP})'v_{0}-w_{1}(v_{0\cP})'+(w_{0\cP})'v_{1-}
 -w_{0}(v_{1-})'-w_{0}v_{0\cP}-w_{0\cP}v_{0},
\label{eq:c+6}
\end{align}
and for the lower sign if and only if
\begin{align}
&4V_{0}^{-}=-w'_{1}-2w_{0}+(w_{1})^{2}+(v_{1-})^{2}-4C_{00},
\label{eq:c-1}\\
&4V_{1}^{-}=(v_{1-})'-2v_{0+}-2w_{1-}v_{1-}-4C_{01},
\label{eq:c-2}\\
&4V_{0}^{-\prime}=-w''_{1}-2w'_{0}+2w_{1}w'_{1}+2v_{1-}(v_{1-})',
\label{eq:c-3}\\
&4V_{1}^{-\prime}=(v_{1-})''-2v'_{0}-2(w_{1-})'v_{1-}-2w_{1-}(v_{1-})'
 +w_{1}v_{0}+w_{1\cP}v_{0\cP}+2w_{0-}v_{1-},
\label{eq:c-4}\\
&2V_{0}^{-\prime\prime}-4(V_{0}^{-}+C_{00})^{2}-4(V_{1}^{-}+C_{01})\left(
 (V_{1}^{-})_{\cP}+C_{01}\right) -4C_{10}=\notag\\
&\hspace*{30pt}-w''_{0}+w'_{1}w_{0}+w_{1}w'_{0}-(v_{1-})'v_{0\cP}-v_{1-}(v_{0\cP})'
 -(w_{0})^{2}-(v_{0\cP})^{2},
\label{eq:c-5}\\
&2V_{1}^{-\prime\prime}-8\left( (V_{0}^{-})_{+}+C_{00}\right) (V_{1}^{-}+C_{01})
 -4C_{11}=-v''_{0}+w'_{1}v_{0}+w_{1}v'_{0}\notag\\
&\hspace*{30pt}-(w_{0\cP})'v_{1-}-w_{0\cP}(v_{1-})'-w_{0}v_{0}-w_{0\cP}v_{0\cP}.
\label{eq:c-6}
\end{align}
The formulas (\ref{eq:c+1}), (\ref{eq:c+2}), (\ref{eq:c-1}), and (\ref{eq:c-2}) determine
the form of all the potential terms $V_{0}^{\pm}$ and $V_{1}^{\pm}$. In addition, they
are automatically compatible with (\ref{eq:co1})--(\ref{eq:co3}), (\ref{eq:c+3}), and
(\ref{eq:c-3}). From (\ref{eq:c+2}) and (\ref{eq:c-2}), we see that both the potential
terms $V_{1}^{+}(q)$ and $V_{1}^{-}(q)$ are even and thus we do not need to check
the other intertwining relation $P_{2}^{+}H^{+}-H^{-}P_{2}^{+}=0$ in (\ref{eq:NS1}).
Hence, there remain nine conditions, (\ref{eq:co4})--(\ref{eq:co6}),
(\ref{eq:c+4})--(\ref{eq:c+6}), and (\ref{eq:c-4})--(\ref{eq:c-6}) to be investigated.
Let us first begin with (\ref{eq:co4}), (\ref{eq:c+4}), and (\ref{eq:c-4}). By the
substitution of (\ref{eq:c+1}), (\ref{eq:c+2}), (\ref{eq:c-1}), and (\ref{eq:c-2}) into
them, they read as
\begin{align}
&2(v_{0-})'+(w_{1+})'v_{1-}-w_{1+}(v_{1-})'-2w_{1+}v_{0+}-2w_{0-}v_{1-}
 -4C_{01}w_{1+}=0,
\label{eq:co4'}\\
&(v_{0-})'+(w_{1+})'v_{1-}-w_{1+}(v_{1-})'-w_{1+}v_{0+}+w_{1-}v_{0-}-w_{0-}v_{1-}=0,
\label{eq:c+4'}\\
&(v_{0-})'-w_{1+}v_{0+}-w_{1-}v_{0-}-w_{0-}v_{1-}=0.
\label{eq:c-4'}
\end{align}
It is easy to check that they are compatible with each other if and only if
\begin{align}
C_{01}w_{1+}=0.
\label{eq:w1+}
\end{align}
{}From (\ref{eq:c+4'}) and (\ref{eq:c-4'}), we obtain
\begin{align}
(w_{1+})'v_{1-}-w_{1+}(v_{1-})'+2w_{1-}v_{0-}=0.
\label{eq:co7}
\end{align}
Hence, the conditions (\ref{eq:co4}), (\ref{eq:c+4}), and (\ref{eq:c-4}) are equivalent
to (\ref{eq:c-4'})--(\ref{eq:co7}). We note in particular that Eq.~(\ref{eq:co7}) enables
us to express $w_{1-}$ (or $v_{0-}$) in terms of $w_{1+}$, $v_{1-}$, and $v_{0-}$
(or $w_{1-}$), respectively. Next, we shall investigate (\ref{eq:co5}),
(\ref{eq:c+5}), and (\ref{eq:c-5}). Substituting (\ref{eq:c+1}), (\ref{eq:c+2}),
(\ref{eq:c-1}), and (\ref{eq:c-2}) into them, we have the three conditions
(\ref{eq:co5'})--(\ref{eq:c-5'}) presented in Appendix. We can easily check that
they are equivalent to the following set of conditions:
\begin{align}
&2w_{1}w''_{1}-(w'_{1})^{2}+2v_{1-}(v_{1-})''-((v_{1-})')^{2}-4v_{1-}(v_{0-})'
 +4(v_{0-})^{2}-2(w_{1})^{2}w'_{1}\notag\\
&+4(w_{1})^{2}w_{0}-2w'_{1}(v_{1-})^{2}-4w_{1-}v_{1-}(v_{1-})'-8w_{1-}v_{1-}v_{0+}
 +4w_{0}(v_{1-})^{2}\notag\\
&-(w_{1})^{4}-2\left[(w_{1})^{2}+2(w_{1-})^{2}\right] (v_{1-})^{2}-(v_{1-})^{4}
 -16C_{10}=0,
\label{eq:co8}\\
&w'''_{1}-w_{1}w''_{1}-2(w'_{1})^{2}+4w'_{1}w_{0}+2w_{1}w'_{0}-v_{1-}(v_{1-})''
 -2((v_{1-})')^{2}\notag\\
&-2(v_{1-})'(2v_{0+}-v_{0-})-2v_{1-}(v_{0+})'+4v_{0+}v_{0-}-2(w_{1})^{2}w'_{1}
 \notag\\
&-2w'_{1}(v_{1-})^{2}-4w_{1-}v_{1-}(v_{1-})'=0,
\label{eq:co8'}\\
&C_{01}v_{0-}=0,
\label{eq:v0-}
\end{align}
where Eq.~(\ref{eq:co7}) has been applied for the derivation of the last formula.
Now, the remaining conditions to be examined are (\ref{eq:co6}), (\ref{eq:c+6}),
and (\ref{eq:c-6}). Substituting (\ref{eq:c+1}), (\ref{eq:c+2}),
(\ref{eq:c-1}), and (\ref{eq:c-2}) into them, we have the three conditions
(\ref{eq:co6'})--(\ref{eq:c-6'}) presented in Appendix. We can easily check that
they are equivalent to the following set of conditions:
\begin{align}
&w''_{1}v_{1-}-(w_{1-})'(v_{1-})'-w_{1\cP}(v_{1-})''+2(w_{1-})'v_{0-}+2w_{1}(v_{0-})'
 -2w_{1-}(w_{1-})'v_{1-}\notag\\
&-\left[(w_{1+})^{2}+(w_{1-})^{2}\right] (v_{1-})'-2\left[ (w_{1+})^{2}+(w_{1-})^{2}
 \right] v_{0+}+4w_{1-}w_{0+}v_{1-}-(v_{1-})^{2}(v_{1-})'\notag\\
&-2(v_{1-})^{2}v_{0+}-2\left[(w_{1+})^{2}+(w_{1-})^{2}\right]w_{1-}v_{1-}
 -2w_{1-}(v_{1-})^{3}+8C_{11}=0,
\label{eq:co9}\\
&(v_{1-})'''+2(v_{0-})''+(w_{1\cP})''v_{1-}-4(w_{1-})'(v_{1-})'-w_{1}(v_{1-})''
 -2(w_{1+})'v_{0}\notag\\
&-4(w_{1-})'v_{0+}-2w_{1}(v_{0+})'+2(w_{0\cP})'v_{1-}+2(2w_{0+}-w_{0-})(v_{1-})'
 +4w_{0-}v_{0-}\notag\\
&-4w_{1-}(w_{1-})'v_{1-}-2\left[(w_{1+})^{2}+(w_{1-})^{2}\right] (v_{1-})'
 -2(v_{1-})^{2}(v_{1-})'=0,
\label{eq:co9'}\\
&C_{01}w_{0-}=0,
\label{eq:w0-}
\end{align}
where Eqs.~(\ref{eq:c-4'}) and (\ref{eq:co7}) have been applied for the derivation
of the last formula. From (\ref{eq:w1+}), (\ref{eq:v0-}), and (\ref{eq:w0-}), we
conclude that
\begin{align}
C_{01}=0\qquad\text{or}\qquad w_{1+}=v_{0-}=w_{0-}=0.
\label{eq:c10}
\end{align}
To analyze (\ref{eq:co8}), (\ref{eq:co8'}), (\ref{eq:co9}), and (\ref{eq:co9'}), we first
note that the equalities must hold for their even and odd parts separately since we
obtain another set of equalities by applying a reflection operator to them.
The even and odd parts of them are explicitly presented in Appendix,
(\ref{eq:co8+})--(\ref{eq:co9'-}). It is apparent that (\ref{eq:co9-}) and
(\ref{eq:co9'-}) are automatically satisfied under the conditions (\ref{eq:c-4'})
and (\ref{eq:co7}). Hence, the remaining problem is to solve
(\ref{eq:co8+})--(\ref{eq:co9+}) and (\ref{eq:co9'+}) simultaneously. However,
they are not independent under the conditions (\ref{eq:c-4'}) and (\ref{eq:co7}).
In fact, we can check that the following combinations
\begin{align*}
&2w_{1-}\times\text{(\ref{eq:co8'+})}+2w_{1+}\times\text{(\ref{eq:co8'-})}
 +2v_{1}\times\text{(\ref{eq:co9'+})}+4v_{0}\times[2\times\text{(\ref{eq:c-4'})}
 +\text{(\ref{eq:co7})}],\\
&w_{1+}\times\text{(\ref{eq:co8'+})}+w_{1-}\times\text{(\ref{eq:co8'-})}
 -(v'_{1}+2v_{0+})\times\text{(\ref{eq:co7})},\\
&v_{1}\times\text{(\ref{eq:co8'+})}+w_{1-}\times\text{(\ref{eq:co9'+})}
 +[(w_{1+})'-2w_{0-}]\times\text{(\ref{eq:co7})},
\end{align*}
are identical with the equations obtained by differentiating (\ref{eq:co8+}),
(\ref{eq:co8-}), and (\ref{eq:co9+}), respectively. In other words, the set of
equations (\ref{eq:co8+}), (\ref{eq:co8-}), and (\ref{eq:co9+}) are equivalent
to the set of equations (\ref{eq:co8'+}), (\ref{eq:co8'-}), and (\ref{eq:co9'+})
under the conditions (\ref{eq:c-4'}) and (\ref{eq:co7}). Therefore, the remaining
task we should settle is now to solve only the former with (\ref{eq:c-4'}),
(\ref{eq:co7}), and (\ref{eq:c10}). In what follows,
we shall analyze separately the two cases of $C_{01}=0$ and $C_{01}\neq 0$.

\subsection{The $C_{01}=0$ Case}

In this case, the condition (\ref{eq:c10}) does not provide any constraint on the three
functions $w_{1+}$, $v_{0-}$, and $w_{0-}$. Hence, all that we should do is to solve
(\ref{eq:co8+}), (\ref{eq:co8-}), and (\ref{eq:co9+}) under the two conditions
(\ref{eq:c-4'}) and (\ref{eq:co7}). It is actually easy since they can be regarded as
simultaneous linear equations for $w_{0+}$, $w_{0-}$, and $v_{0+}$:
\begin{align}
2A\left( \begin{array}{c}w_{0+}\\ w_{0-}\\ v_{0+}\\ \end{array}\right)=
 \left(\begin{array}{c}f_{1}\\ f_{2}\\ f_{3}\\ \end{array}\right),
\end{align}
where $f_{i}$ ($i=1,2,3$) which only depend on $w_{1+}$, $w_{1-}$, and $v_{1-}$ are
explicitly presented in (\ref{eq:f1})--(\ref{eq:f3}) while the $3\times 3$ matrix $A$ is
given by
\begin{align}
A=\left(
 \begin{array}{ccc}
 2\left[(w_{1+})^{2}+(w_{1-})^{2}+(v_{1-})^{2}\right] & 4w_{1+}w_{1-} & -4w_{1-}v_{1-}\\
 2w_{1+}w_{1-} & (w_{1+})^{2}+(w_{1-})^{2} & -w_{1+}v_{1-}\\
 2w_{1-}v_{1-} & w_{1+}v_{1-} & -(w_{1-})^{2}-(v_{1-})^{2}\\
 \end{array}\right).
\end{align}
Hence, we must treat the problem separately according to the value of the
determinant of $A$:
\begin{align}
\det A=-2(w_{1-})^{2}\left[ (w_{1+})^{2}-(w_{1-})^{2}+(v_{1-})^{2}\right]^{2}.
\end{align}

\noindent
\textbf{Case 1.} $(w_{1-})^{2}\neq (w_{1+})^{2}+(v_{1-})^{2}$ and $w_{1-}\neq0$:\\

In the non-degenerate case $\det A\neq 0$, they are uniquely solved as
\begin{align}
&-2(\det A)\,w_{0+}=(w_{1-})^{2}\left[(w_{1+})^{2}+(w_{1-})^{2}+(v_{1-})^{2}\right]
 f_{1}-4w_{1+}(w_{1-})^{3}f_{2}\notag\\
&\qquad-4(w_{1-})^{3}v_{1-}f_{3},
\label{eq:w0+}\\
&-(\det A)\,w_{0-}=-w_{1+}(w_{1-})^{3}f_{1}+\left[(w_{1+})^{2}(w_{1-})^{2}+(w_{1-})^{4}
 +(w_{1+})^{2}(v_{1-})^{2}\right.\notag\\
&\qquad\left.-2(w_{1-})^{2}(v_{1-})^{2}+(v_{1-})^{4}\right]f_{2}-\left[ (w_{1+})^{2}
 -3(w_{1-})^{2}+(v_{1-})^{2}\right]w_{1+}v_{1-}f_{3},
\\
&-(\det A)\,v_{0+}=(w_{1-})^{3}v_{1-}f_{1}+\left[(w_{1+})^{2}-3(w_{1-})^{2}+(v_{1-})^{2}
 \right]w_{1+}v_{1-}f_{2}\notag\\
&\qquad -\left[(w_{1+})^{4}-2(w_{1+})^{2}(w_{1-})^{2}
 +(w_{1-})^{4}+(w_{1+})^{2}(v_{1-})^{2}+(w_{1-})^{2}(v_{1-})^{2}\right]f_{3},
\label{eq:v0+}
\end{align}
Therefore, the most general $2$-fold SUSY quantum systems with reflections
composed of $H^{\pm}$ in (\ref{eq:H+-}) and $P_{2}^{\pm}$ in (\ref{eqs:P2+-})
are entirely expressible solely in terms of three arbitrary functions having
definite parity $w_{1+}$, $w_{1-}$, and $v_{1-}$ by using (\ref{eq:co7}) and
(\ref{eq:w0+})--(\ref{eq:v0+}) in the non-degenerate case. When $v_{1-}=v_{0+}=0$,
then $v_{0-}=0$ from (\ref{eq:co7}) and the set of identities
(\ref{eq:w0+})--(\ref{eq:v0+}) reduces to
\begin{align}
4(w_{1})^{2}w_{0}=-2w_{1}(w_{1})''+((w_{1})')^{2}+2(w_{1})^{2}(w_{1})'+(w_{1})^{4}
 +16C_{10},\quad C_{11}=0.
\end{align}
Hence, the systems in this case reduces to the most general $2$-fold SUSY ordinary
quantum systems without reflections in Refs.~\cite{AST01b,AICD95,AIN95b}.\\

\noindent
\textbf{Case 2.} $(w_{1-})^{2}=(w_{1+})^{2}+(v_{1-})^{2}\neq 0$:\\

Next, we shall examine the degenerate case 
\begin{align}
(w_{1-})^{2}=(w_{1+})^{2}+(v_{1-})^{2}\neq 0.
\label{eq:c2f1}
\end{align}
In this case, the three equations (\ref{eq:co8+}), (\ref{eq:co8-}), and (\ref{eq:co9+})
are not linearly independent and are equivalent to the two equations
\begin{subequations}
\label{eqs:case2}
\begin{align}
8(w_{1-})^{2}(w_{1+}w_{0+}+w_{1-}w_{0-})&=-w_{1+}f_{1}+4w_{1-}f_{2},\\
8(w_{1-})^{2}(v_{1-}w_{0+}-w_{1-}v_{0+})&=-v_{1-}f_{1}+4w_{1-}f_{3},
\end{align}
\end{subequations}
with the constraint
\begin{align}
w_{1-}f_{1}-2w_{1+}f_{2}-2v_{1-}f_{3}=16(C_{10}w_{1-}+C_{11}v_{1-})=0.
\label{eq:c2cst}
\end{align}
We note that we have the following interesting formula by using (\ref{eq:co7}) and
(\ref{eq:c2f1}):
\begin{align}
4(v_{0-})^{2}=((w_{1+})')^{2}-((w_{1-})')^{2}+((v_{1-})')^{2}.
\label{eq:c2f2}
\end{align}
We first show that $C_{10}=C_{11}=0$ to satisfy the constraint (\ref{eq:c2cst}). Suppose
$C_{11}\neq 0$ since $C_{11}=0$ inevitably means $C_{10}=0$ due to the assumption
$w_{1-}\neq 0$ in this case. Then, we have $v_{1-}=-C_{10}w_{1-}/C_{11}:=\tilde{C}_{1}
w_{1-}$ from (\ref{eq:c2cst}). Substituting it into the current assumption $(w_{1-})^{2}=
(w_{1+})^{2}+(v_{1-})^{2}$, we obtain $w_{1+}=\tilde{C}_{2}w_{1-}$ with $(\tilde{C}_{1}
)^{2}+(\tilde{C}_{2})^{2}=1$. But $w_{1+}$ and $w_{1-}$ are even and odd analytic
functions, respectively, and thus we must conclude that $w_{1+}=w_{1-}=0$, which
contradicts with the assumption $w_{1-}\neq 0$. Hence, we eventually have
$C_{10}=C_{11}=0$.

Finally, using (\ref{eq:c2f1}), (\ref{eqs:case2}), and (\ref{eq:c2f2}), we can eliminate
four functions, e.g., $w_{1-}$, $w_{0-}$, $v_{0+}$, and $v_{0-}$, to express the most
general $2$-fold SUSY quantum systems with reflections in this case in terms of the
remaining three arbitrary functions having definite parity, e.g., $w_{1+}$, $v_{1-}$,
and $w_{0+}$.

We note that the systems in this case do not admit a reduction to ordinary
quantum systems without reflections. Indeed, if we put $v_{1-}=0$, then $v_{0-}=0$
from (\ref{eq:co7}) since $w_{1-}\neq 0$ by the assumption, and thus
$(w_{1+})'=\pm(w_{1-})'$ from (\ref{eq:c2f2}). But $(w_{1+})'$ and $(w_{1-})'$
are odd and even analytic functions, and thus it is inevitable that
$(w_{1+})'=(w_{1-})'=0$. As any non-zero constant is an even function, we must
conclude that $w_{1-}=0$, which contradicts the assumption $w_{1-}\neq0$.
Hence, the systems in this case have no reductions to ordinary
quantum systems without reflections.\\

\noindent
\textbf{Case 3:} $w_{1-}=0$:\\

Next, we shall examine the other degenerate case $w_{1-}=0$. In this case,
the potential terms $V_{0}^{\pm}$, $V_{1}^{\pm}$ and the $2$-fold supercharge
component $P_{2}^{-}$ read from (\ref{eq:c+1}), (\ref{eq:c+2}), (\ref{eq:c-1}),
(\ref{eq:c-2}), and (\ref{eqs:P2+-}) as
\begin{subequations}
\label{eqs:de2}
\begin{align}
4V_{0}^{+}&=3(w_{1+})'-2w_{0+}-2w_{0-}+(w_{1+})^{2}+(v_{1-})^{2}-4C_{00},\\
4V_{1}^{+}&=-3(v_{1-})'-2v_{0+},\\
4V_{0}^{-}&=-(w_{1+})'-2w_{0+}-2w_{0-}+(w_{1+})^{2}+(v_{1-})^{2}-4C_{00},\\
4V_{1}^{-}&=(v_{1-})'-2v_{0+},\\
P_{2}^{-}&=\frac{\rmd^{2}}{\rmd q^{2}}+(w_{1+}+v_{1-}\cP)\frac{\rmd}{\rmd q}
 +w_{0+}+w_{0-}+(v_{0+}+v_{0-})\cP.
\end{align}
\end{subequations}
The conditions (\ref{eq:c-4'}), (\ref{eq:co7}), (\ref{eq:co8+}), (\ref{eq:co8-}), and
(\ref{eq:co9+}), which are all that we must manage, read as
\begin{subequations}
\label{eqs:d2c}
\begin{align}
&(v_{0-})'-w_{1+}v_{0+}-w_{0-}v_{1-}=0,\\
&(w_{1+})'v_{1-}-w_{1+}(v_{1-})'=0,\\
&4\left[ (w_{1+})^{2}+(v_{1-})^{2}\right] w_{0+}=-2w_{1+}(w_{1+})''+((w_{1+})')^{2}
 -2v_{1-}(v_{1-})''\notag\\
&+((v_{1-})')^{2}-4(v_{0-})^{2}+(w_{1+})^{4}+2(w_{1+})^{2}(v_{1-})^{2}+(v_{1-})^{4}
 +16C_{10},\\
&2(w_{1+})^{2}w_{0-}-2w_{1+}v_{1-}v_{0+}=(w_{1+})^{2}(w_{1+})'+w_{1+}v_{1-}
 (v_{1-})',\\
&2w_{1+}v_{1-}w_{0-}-2(v_{1-})^{2}v_{0+}=w_{1+}(w_{1+})'v_{1-}+(v_{1-})^{2}
 (v_{1-})'-8C_{11}.
\end{align}
\end{subequations}
The second equality means that the two functions $w_{1+}$ and $v_{1-}$ are
linearly dependent unless at least either of them vanishes. But they cannot
be linearly dependent since $w_{1+}$ and $v_{1-}$ are even and odd analytic
functions, respectively. Hence, we conclude that $w_{1+}v_{1-}=0$. It turns out
that we have three inequivalent solutions to them as the followings.\\

\noindent
\textbf{Case 3-1.} $w_{1+}\neq 0$ and $v_{1-}=0$:\\

In this case, the set of the conditions (\ref{eqs:d2c}) are solved as
\begin{align}
\begin{split}
&w_{0+}=-\frac{(w_{1+})''}{2w_{1+}}+\frac{((w_{1+})')^{2}}{4(w_{1+})^{2}}
 -\frac{(v_{0-})^{2}}{(w_{1+})^{2}}+\frac{(w_{1+})^{2}}{4}+\frac{4C_{10}}{(w_{1+})^{2}},\\
&w_{0-}=\frac{(w_{1+})'}{2},\qquad v_{0+}=\frac{(v_{0-})'}{w_{1+}},\qquad C_{11}=0.
\end{split}
\label{eq:d2s1}
\end{align}
Hence, the most general $2$-fold SUSY quantum systems with reflections in this
case are expressed in terms of the two arbitrary functions having definite parity
$w_{1+}$ and $v_{0-}$ by the substitution of (\ref{eq:d2s1}) into (\ref{eqs:de2}).
When $v_{0-}=0$, it is inevitable that $v_{0+}=0$ from (\ref{eq:d2s1}). Then, we have
\begin{align}
w_{0}=\frac{(w_{1+})^{2}}{4}-\frac{(w_{1+})''}{2w_{1+}}
 +\frac{((w_{1+})')^{2}}{4(w_{1+})^{2}}+\frac{(w_{1+})^{2}}{4}
 +\frac{4C_{10}}{(w_{1+})^{2}}+\frac{(w_{1+})'}{2},
\label{eq:w0c31}
\end{align}
and the systems (\ref{eqs:de2}) with (\ref{eq:w0c31}) exactly reduce to
the general $2$-fold SUSY ordinary quantum systems without reflections in
Refs.~\cite{AST01b,AICD95,AIN95b} but with the restriction $w_{1}=w_{1+}$.\\

\noindent
\textbf{Case 3-2.} $w_{1+}=0$ and $v_{1-}\neq 0$:\\

In this case, the set of the conditions (\ref{eqs:d2c}) are solved as
\begin{align}
\begin{split}
&w_{0+}=-\frac{(v_{1-})''}{2v_{1-}}+\frac{((v_{1-})')^{2}}{4(v_{1-})^{2}}
 -\frac{(v_{0-})^{2}}{(v_{1-})^{2}}+\frac{(v_{1-})^{2}}{4}+\frac{4C_{10}}{(v_{1-})^{2}},\\
&w_{0-}=\frac{(v_{0-})'}{v_{1-}},\qquad v_{0+}=-\frac{(v_{1-})'}{2}
 +\frac{4C_{11}}{(v_{1-})^{2}}.
\end{split}
\label{eq:d2s2}
\end{align}
Hence, the most general $2$-fold SUSY quantum systems with reflections in this
case are expressed in terms of the two arbitrary functions having definite parity
$v_{1-}$ and $v_{0-}$ by the substitution of (\ref{eq:d2s2}) into (\ref{eqs:de2}).
In contrast to the previous Case 3-1, the assumption $v_{1-}\neq 0$ does not
admit of a reduction of the systems to ordinary quantum systems without
reflections.\\

\noindent
\textbf{Case 3-3.} $w_{1+}=v_{1-}=0$:\\

In this case, the set of the conditions (\ref{eqs:d2c}) are solved as
\begin{align}
(v_{0-})^{2}=4C_{10},\qquad C_{11}=0.
\label{eq:d2s3}
\end{align}
Hence, the most general $2$-fold SUSY quantum systems with reflections in this
case are expressed in terms of the three arbitrary functions having definite parity
$w_{0+}$, $w_{0-}$, and $v_{0+}$ by the substitution of (\ref{eq:d2s3}) into
(\ref{eqs:de2}). This case is almost trivial since we have from (\ref{eqs:de2})
\begin{align}
\begin{split}
&2V_{0}^{+}=2V_{0}^{-}=-w_{0}-2C_{00},\quad 2V_{1}^{+}=2V_{1}^{-}=-v_{0+},\\
&P_{2}^{-}=\frac{\rmd^{2}}{\rmd q^{2}}+w_{0}+\left( v_{0+}\pm 2\sqrt{C_{10}}
 \right)\cP,
\end{split}
\end{align}
and thus the pair of $2$-fold SUSY Hamiltonians coincides $H^{+}=H^{-}$. They
reduce to the corresponding trivial $2$-fold SUSY ordinary quantum systems
without reflections when $v_{0+}=v_{0-}=C_{10}=0$.

\subsection{The $C_{01}\neq 0$ Case}

In this case, it is inevitable from (\ref{eq:c10}) that
\begin{align}
w_{1+}=v_{0-}=w_{0-}=0,
\label{eq:wvw}
\end{align}
and the potential terms $V_{0}^{\pm}$, $V_{1}^{\pm}$ and the $2$-fold supercharge
component $P_{2}^{-}$ read from (\ref{eq:c+1}), (\ref{eq:c+2}), (\ref{eq:c-1}),
(\ref{eq:c-2}), and (\ref{eqs:P2+-}) as
\begin{subequations}
\label{eqs:ne0}
\begin{align}
4V_{0}^{+}&=3(w_{1-})'-2w_{0+}+(w_{1-})^{2}+(v_{1-})^{2}-4C_{00},\\
4V_{1}^{+}&=-3(v_{1-})'-2v_{0+}-2w_{1-}v_{1-}-4C_{01},\\
4V_{0}^{-}&=-(w_{1-})'-2w_{0+}+(w_{1-})^{2}+(v_{1-})^{2}-4C_{00},\\
4V_{1}^{-}&=(v_{1-})'-2v_{0+}-2w_{1-}v_{1-}-4C_{01},\\
P_{2}^{-}&=\frac{\rmd^{2}}{\rmd q^{2}}+(w_{1-}+v_{1-}\cP)\frac{\rmd}{\rmd q}
 +w_{0+}+v_{0+}\cP.
\end{align}
\end{subequations}
We note that all the potential terms $V_{0}^{\pm}(q)$ and $V_{1}^{\pm}(q)$ have
even parity.
The conditions (\ref{eq:c-4'}), (\ref{eq:co7}), and (\ref{eq:co8-}) are automatically
satisfied, and thus all the remaining conditions to be solved are (\ref{eq:co8+}) and
(\ref{eq:co9+}) which now read as
\begin{align}
2B\left( \begin{array}{c}w_{0+}\\ v_{0+}\\ \end{array}\right)=
 \left( \begin{array}{c}f_{1}\\ f_{3}\\ \end{array}\right),\qquad
 B=\left(\begin{array}{cc}
 2\left[ (w_{1-})^{2}+(v_{1-})^{2}\right] & -4w_{1-}v_{1-}\\
 2w_{1-}v_{1-} & -(w_{1-})^{2}-(v_{1-})^{2}\\ \end{array}\right),
\end{align}
where $f_{1}$ and $f_{3}$ are now given by
\begin{subequations}
\label{eqs:n0c}
\begin{align}
&f_{1}=-2w_{1-}(w_{1-})''+((w_{1-})')^{2}-2v_{1-}(v_{1-})''+((v_{1-})')^{2}
 +2(w_{1-})^{2}(w_{1-})'\notag\\
&+2(w_{1-})'(v_{1-})^{2}+4w_{1-}v_{1-}(v_{1-})'+(w_{1-})^{4}+6(w_{1-})^{2}
 (v_{1-})^{2}+(v_{1-})^{4}+16C_{10},\\
&f_{3}=-(w_{1-})''v_{1-}+(w_{1-})'(v_{1-})'-w_{1-}(v_{1-})''+2w_{1-}(w_{1-})'v_{1-}
 +(w_{1-})^{2}(v_{1-})'\notag\\
&+(v_{1-})^{2}(v_{1-})'+2(w_{1-})^{3}v_{1-}+2w_{1-}(v_{1-})^{3}-8C_{11}.
\end{align}
\end{subequations}
They are simultaneous linear equations for $w_{0+}$ and $v_{0+}$. Hence, we must
treat the problem according to the value of $\det B=-2[(w_{1-})^{2}-(v_{1-})^{2}]^{2}$.\\

\noindent
\textbf{Case 4.} $(v_{1-})^{2}\neq (w_{1-})^{2}$:\\

In the non-degenerate case $(v_{1-})^{2}\neq (w_{1-})^{2}$, they are uniquely solved as
\begin{align}
\begin{split}
-2(\det B)w_{0+}&=\left[(w_{1-})^{2}+(v_{1-})^{2}\right] f_{1}-4w_{1-}v_{1-}f_{3},\\
-(\det B)v_{0+}&=w_{1-}v_{1-}f_{1}-\left[(w_{1-})^{2}+(v_{1-})^{2}\right] f_{3}.
\end{split}
\label{eq:case4}
\end{align}
Hence, the most general $2$-fold SUSY quantum systems with reflections in this
case are expressed in terms of the two arbitrary functions having definite parity
$w_{1-}$, and $v_{1-}$ by the substitution of (\ref{eq:case4}) for $w_{0+}$ and
$v_{0+}$ into (\ref{eqs:ne0}). Due to the assumption $C_{10}\neq 0$, the systems
do not admit a reduction to ordinary quantum systems without reflections. When
we put $C_{01}=0$, the systems in this case reduces to the ones in Case 1 with the
constraint (\ref{eq:wvw}).\\

\noindent
\textbf{Case 5.} $v_{1-}=\pm w_{1-}\neq 0$:\\

In the degenerate case $v_{1-}=\pm w_{1-}\neq 0$, on the other hand, the two
equations in (\ref{eqs:n0c}) are not linearly independent and are equivalent to
the following single equation:
\begin{align}
4(w_{1-})^{2}(w_{0+}\mp v_{0+})=-2w_{1-}(w_{1-})''+((w_{1-})')^{2}+4(w_{1-})^{2}
 (w_{1-})'+4(w_{1-})^{4}+8C_{10},
\end{align}
with $C_{11}=\mp C_{10}$. Hence, we can again express the most general $2$-fold
SUSY quantum systems with reflections in terms of two functions having definite
parity, e.g., $w_{1-}$ and $w_{0+}$, by eliminating the other two functions.
Due to the assumption $C_{10}\neq 0$, the systems do not admit a reduction to
ordinary quantum systems without reflections. When we put $C_{01}=0$, the systems
in this case reduces to the ones in Case 2 with the constraint (\ref{eq:wvw}).\\

It is worth noting that the $2$-fold SUSY quantum systems with reflections
characterized by the constraint (\ref{eq:wvw}) are essentially equivalent
to the $2\times 2$ Hermitian matrix $2$-fold SUSY quantum systems in
Ref.~\cite{Ta12a}. To see the relation, we first split the linear function space
$\fF$ in which the systems have been considered into its even and odd parts,
denoted by $\fF_{+}$ and $\fF_{-}$, respectively. Their elements
$\psi_{+}(q)\in\fF_{+}$ and $\psi_{-}(q)\in\fF_{-}$ are even and odd functions,
respectively, and thus $\cP\cdot\psi_{\pm}(q)=\pm\psi_{\pm}(q)$. With this
grading, we can introduce a two-component representation of $\psi\in\fF$
as~\cite{PVZ11}
\begin{align}
\psi(q)\stackrel{\text{rep.}}{=}\left( \begin{array}{r}\psi_{+}(q)\\ \psi_{-}(q)\\
 \end{array}\right).
\end{align}
Then, a reflection operator $\cP$, a differential operator $\rmd/\rmd q$,
and any multiplicative operator of even and odd functions, $f_{+}(q)$ and
$f_{-}(q)$, are represented by $2\times 2$ matrices as
\begin{align}
\begin{split}
&\cP\stackrel{\text{rep.}}{=}\left( \begin{array}{cc} 1 & 0\\ 0 & -1\\ \end{array}
 \right)=\sigma_{3},\quad \frac{\rmd}{\rmd q}\stackrel{\text{rep.}}{=}\left(
 \begin{array}{cc} 0 & \rmd/\rmd q\\ \rmd/\rmd q & 0\\ 
 \end{array}\right)=\sigma_{1}\frac{\rmd}{\rmd q},\\
&f_{+}\stackrel{\text{rep.}}{=}\left( \begin{array}{cc} f_{+} & 0\\ 0 & f_{+}\\
 \end{array}\right)=f_{+}I_{2},\quad f_{-}\stackrel{\text{rep.}}{=}\left(
 \begin{array}{cc} 0 & f_{-}\\ f_{-} & 0\\ \end{array}\right)=f_{-}\sigma_{1},
\end{split}
\end{align}
where $\sigma_{1}$ and $\sigma_{3}$ are the Pauli matrices and $I_{2}$ is the
$2\times 2$ unit matrix. In this representation, the $2$-fold SUSY systems
$H^{\pm}$ and $P_{2}^{-}$ in the present case (\ref{eqs:ne0}) are expressed as
\begin{subequations}
\label{eqs:2rep}
\begin{align}
H^{+}\stackrel{\text{rep.}}{=}&\;-\frac{1}{2}I_{2}\frac{\rmd^{2}}{\rmd q^{2}}
 +\frac{1}{4}\left[3(w_{1-})'
 -2w_{0+}+(w_{1-})^{2}+(v_{1-})^{2}-4C_{00}\right] I_{2}\notag\\
&\;-\frac{1}{4}\left[3(v_{1-})'+2v_{0+}+2w_{1-}v_{1-}-4C_{01}\right]\sigma_{3},\\
H^{-}\stackrel{\text{rep.}}{=}&\;-\frac{1}{2}I_{2}\frac{\rmd^{2}}{\rmd q^{2}}
 +\frac{1}{4}\left[-(w_{1-})'
 -2w_{0+}+(w_{1-})^{2}+(v_{1-})^{2}-4C_{00}\right] I_{2}\notag\\
&\;-\frac{1}{4}\left[-(v_{1-})'+2v_{0+}+2w_{1-}v_{1-}-4C_{01}\right]\sigma_{3},\\
P_{2}^{-}\stackrel{\text{rep.}}{=}&\;I_{2}\frac{\rmd^{2}}{\rmd q^{2}}
 +\left(w_{1-}I_{2}-v_{1-}\sigma_{3}
 \right)\frac{\rmd}{\rmd q}+w_{0+}I_{2}+v_{0+}\sigma_{3}.
\end{align}
\end{subequations}
Comparing the above with the most general $2\times 2$ Hermitian matrix $2$-fold
SUSY systems in Ref.~\cite{Ta12a}, we find that the above $2\times 2$ matrix
$2$-fold SUSY systems (\ref{eqs:2rep}) are identical to the latter with the following
substitutions:
\begin{align}
\begin{split}
&w_{10}\to w_{1-},\quad v_{1}\to v_{1-},\quad w_{00}\to w_{0+},\quad v_{0}\to
 -v_{0+},\\
&C_{0}\to C_{00},\quad C_{01}, C_{02}\to 0,\quad C_{03}\to -1,\quad\tilde{C}\to
 -C_{11}.
\end{split}
\end{align}
The two conditions (\ref{eqs:n0c}) are also identical with the corresponding ones
for the latter systems (cf., Eqs.~(35) and (36) in Ref.~\cite{Ta12a}) with the above
substitutions. However, there exist differences between them, that is, in our present
 systems all the functions $w_{1-}$, $v_{1-}$, $w_{0+}$, and $v_{0+}$ have definite
parity but are in general complex while in the Hermitian models all the functions
$w_{10}$, $v_{1}$, $w_{00}$, and $v_{0}$ do not have definite parity in general but
are restricted to be real.

\section{Discussion and Summary}
\label{sec:discus}

In this article, we have for the first time formulated generically $\cN$-fold SUSY
in quantum mechanical systems with reflections and constructed the most general
$1$- and $2$-fold SUSY systems. We have found in particular that there are seven
inequivalent cases of $2$-fold SUSY realized by quantum systems with reflections.
In Cases 1, 2, 3-3 they are characterized by three arbitrary functions having definite
parity while in Cases 3-1, 3-2, 4, 5 they are by two. Furthermore, the systems in
Cases 1, 3-1, and 3-3 reduce to the corresponding general $2$-fold SUSY ordinary
quantum systems without reflections while Cases 2, 3-2, 4, and 5 do not. Hence,
it turns out that $\cN$-fold SUSY in quantum systems with reflections has much
richer structure than in ordinary systems without reflections.
In addition to the detailed studies for $\cN>2$ cases, there are many future issues to
be followed after this work as the followings:\\

\noindent
1. The fact that the most general $2$-fold SUSY quantum systems with reflections
 include as particular cases those which are essentially equivalent with $2\times 2$
 Hermitian matrix $2$-fold SUSY quantum models indicates that the former could
 have relation to more general $2\times 2$ non-Hermitian or higher-dimensional
 matrix $2$-fold SUSY quantum models. Or there might exist a unified framework
 of $\cN$-fold SUSY which includes both quantum mechanical systems with
 reflections and matrix models as special cases. In this respect, it is interesting to
 generalize the formulation of $\cN$-fold SUSY to quantum mechanical matrix
 models with reflection operators.\\

\noindent
2. It is important to clarify general aspects of $\cN$-fold SUSY in quantum
 systems with reflections, as were done in~\cite{AST01b,AS03} for ones without
 reflections. In the latter case, there are two significant features, namely, the
 equivalence between $\cN$-fold SUSY and weak quasi-solvability and the
 equivalence between the conditions (\ref{eq:NS1}) and (\ref{eq:NS2}). In the
 case of $2$-fold SUSY quantum systems with reflections investigated in
 Section~\ref{sec:2f}, however, the latter equivalence would be violated since
 the condition (\ref{eq:co4'}) coming from (\ref{eq:NS1}) is evidently weaker
 than the conditions (\ref{eq:c+4'}) and (\ref{eq:c-4'}) coming from (\ref{eq:NS2}).
 That was exactly the reason why we considered the both to derive (\ref{eq:w1+})
 and (\ref{eq:co7}). In this respect, it is interesting to study what happens when
 we employ only one of the conditions (\ref{eq:NS1}) and (\ref{eq:NS2}) exclusively.
 It is also worth applying the general approach for quantum systems without
 reflections recently proposed by us in Ref.~\cite{Ta11a} to ones with reflections.\\

\noindent
3. In the case without reflections, the systematic algorithm for constructing an
 $\cN$-fold SUSY system~\cite{GT05} based on quasi-solvability has shown to
 be quite effective. Hence, its generalization to the present case with reflections
 is desirable. It would enable us to connect directly the possible types of such
 $\cN$-fold SUSY systems with the possible linear spaces of functions preserved
 by a second-order linear differential operator with reflections. It would also help
 us to clarify the structure of the little $-1$ Jacobi polynomials and to obtain
 the family of polynomial systems which arise as a set of exact eigenfunctions
 of an operator of this kind. To the best of our knowledge, there have been no
 systematic investigations into quasi-solvable operators involving reflection
 operators. We would report some results on this subject in our subsequent
 publications.\\

\noindent
4. Shape invariance is a well-known sufficient condition for \emph{solvability}
 of one-dimensional Schr\"{o}dinger equations~\cite{Ge83}. It means in particular
 that it always implies $\cN$-fold SUSY. In fact, some shape-invariant potentials
 in the case without reflections were systematically constructed as particular cases
 of $\cN$-fold SUSY with intermediate Hamiltonians~\cite{BT09,BT10}. To the
 best of knowledge, there have been no investigations into shape-invariant
 potentials with reflections, and we expect that our formulation of $\cN$-fold
 SUSY would be also quite efficient in constructing systematically shape-invariant
 quantum systems with reflections.\\

\noindent
5. Extension to more general second-order linear differential operators with
 reflections would be possible. In particular, a quantum mechanical model with
 reflections having position-dependent mass would be an interesting candidate
 as a natural generalization of $\cN$-fold SUSY in ordinary quantum systems
 with position-dependent mass formulated in Ref.~\cite{Ta06a}.\\

\noindent
6. In the case without reflections, there are several intimate relations between
 $\cN$-fold SUSY and $\cN$th-order paraSUSY~\cite{BT09,BT10,Ta07a,Ta07c}.
 We expect that we can formulate higher-order paraSUSY in quantum systems
 with reflections in a way such that the relations to $\cN$-fold SUSY in the case
 without reflections remain intact in the latter case. Extension of higher-order
 $\cN$-fold paraSUSY~\cite{Ta07b} to quantum systems with reflections would
 be also possible.

%\section*{Acknowledgments}
%\begin{acknowledgments}%REVTEX4
%\ack%IOPART

%\end{acknowledgments}%REVTEX4

\appendix

\section{List of Formulas}

\noindent
The components of anti-commutators of $2$-fold supercharges:
\begin{align}
&P_{2}^{-}P_{2}^{+}=\frac{\rmd^{4}}{\rmd q^{4}}+2v_{1+}\cP\frac{\rmd^{3}}{\rmd q^{3}}
 +\bigl[-3w'_{1}+2w_{0}-(w_{1})^{2}-(v_{1})^{2}+(-3(v_{1\cP})'+2v_{0+}\notag\\
&-w_{1}v_{1\cP}-w_{1\cP}v_{1})\cP\bigr]\frac{\rmd^{2}}{\rmd q^{2}}+\bigl[-3w''_{1}
 +2w'_{0}-2w_{1}w'_{1}-2v_{1}v'_{1}+(3(v_{1\cP})''-2(v_{0\cP})'\notag\\
&+2(w_{1\cP})'v_{1}+2w_{1}(v_{1\cP})'
 -w_{1}v_{0\cP}-w_{1\cP}v_{0}+w_{0}v_{1\cP}+w_{0\cP}v_{1})\cP\bigr]
 \frac{\rmd}{\rmd q}-w'''_{1}+w''_{0}-w_{1}w''_{1}\notag\\[5pt]
&-v_{1}v''_{1}-w'_{1}w_{0}+w_{1}w'_{0}+v'_{1}v_{0}-v_{1}v'_{0}+(w_{0})^{2}
 +(v_{0})^{2}+(-(v_{1\cP})'''+(v_{0\cP})''-(w_{1\cP})''v_{1}\notag\\[10pt]
&-w_{1}(v_{1\cP})''+(w_{1\cP})'v_{0}+w_{1}(v_{0\cP})'-(w_{0\cP})'v_{1}
 -w_{0}(v_{1\cP})'+w_{0}v_{0\cP}+w_{0\cP}v_{0})\cP,
\label{eq:2-2+}
\end{align}
\begin{align}
&P_{2}^{+}P_{2}^{-}=\frac{\rmd^{4}}{\rmd q^{4}}+2v_{1+}\cP\frac{\rmd^{3}}{\rmd q^{3}}
 +\bigl[w'_{1}+2w_{0}-(w_{1})^{2}-(v_{1\cP})^{2}+(-2v'_{1}-(v_{1\cP})'+2v_{0+}\notag\\
&+w_{1}v_{1}+w_{1\cP}v_{1\cP})\cP\bigr]\frac{\rmd^{2}}{\rmd q^{2}}+\bigl[w''_{1}
 +2w'_{0}-2w_{1}w'_{1}-2v_{1\cP}(v_{1\cP})'+(v''_{1}-2v'_{0}-w'_{1}v_{1}\notag\\
&-(w_{1\cP})'v_{1\cP}-w_{1}v'_{1}-w_{1\cP}(v_{1\cP})'+w_{1}v_{0}+w_{1\cP}v_{0\cP}
 +w_{0}v_{1}+w_{0\cP}v_{1\cP})\cP\bigr]\frac{\rmd}{\rmd q}+w''_{0}\notag\\[5pt]
&-w'_{1}w_{0}-w_{1}w'_{0}-(v_{1\cP})'v_{0\cP}-v_{1\cP}(v_{0\cP})'+(w_{0})^{2}
 +(v_{0\cP})^{2}+(v''_{0}-w'_{1}v_{0}-w_{1}v'_{0}\notag\\[10pt]
&-(w_{0\cP})'v_{1\cP}-w_{0\cP}(v_{1\cP})'+w_{0}v_{0}+w_{0\cP}v_{0\cP})\cP.
\label{eq:2+2-}
\end{align}
The three conditions obtained from the substitution of (\ref{eq:c+1}), (\ref{eq:c+2}),
(\ref{eq:c-1}), and (\ref{eq:c-2}) into (\ref{eq:co5}), (\ref{eq:c+5}), and (\ref{eq:c-5}):
\begin{align}
&w'''_{1}-w_{1}w''_{1}-2(w'_{1})^{2}+4w'_{1}w_{0}+2w_{1}w'_{0}-v_{1-}(v_{1-})''
 -2((v_{1-})')^{2}\notag\\
&-2(v_{1-})'(2v_{0+}-v_{0-})-2v_{1-}(v_{0+})'+4v_{0+}v_{0-}-2(w_{1})^{2}w'_{1}
 -2(w_{1-})'(v_{1-})^{2}\notag\\
&-2(w_{1+}+2w_{1-})v_{1-}(v_{1-})'+4w_{1-}v_{1-}v_{0-}+8C_{01}v_{0-}=0,
\label{eq:co5'}\\
&2w'''_{1}-5(w'_{1})^{2}+8w'_{1}w_{0}+4w_{1}w'_{0}-5((v_{1-})')^{2}-4(v_{1-})'
 (2v_{0+}-v_{0-})\notag\\
&-4v_{1-}v'_{0}+4(2v_{0+}+v_{0-})v_{0-}-6(w_{1})^{2}w'_{1}+4(w_{1})^{2}w_{0}
 -6w'_{1}(v_{1-})^{2}\notag\\
&-12w_{1-}v_{1-}(v_{1-})'-8w_{1-}v_{1-}v_{0+}+4w_{0}(v_{1-})^{2}-(w_{1})^{4}
 \notag\\
&-2\left[(w_{1})^{2}+2(w_{1-})^{2}\right] (v_{1-})^{2}-(v_{1-})^{4}-16C_{10}=0,
\label{eq:c+5'}\\
&2w'''_{1}-4w_{1}w''_{1}-3(w'_{1})^{2}+8w'_{1}w_{0}+4w_{1}w'_{0}-4v_{1-}(v_{1-})''
 -3((v_{1-})')^{2}\notag\\
&-4(v_{1-})'(2v_{0+}-v_{0-})-4v_{1-}(v_{0\cP})'+4v_{0-}(2v_{0+}-v_{0-})
 -2(w_{1})^{2}w'_{1}\notag\\
&-4(w_{1})^{2}w_{0}-2w'_{1}(v_{1-})^{2}-4w_{1-}v_{1-}(v_{1-})'+8w_{1-}v_{1-}v_{0+}
 -4w_{0}(v_{1-})^{2}\notag\\
&+(w_{1})^{4}+2\left[(w_{1})^{2}+2(w_{1-})^{2}\right] (v_{1-})^{2}+(v_{1-})^{4}
 +16C_{10}=0.
\label{eq:c-5'}
\end{align}
The three conditions obtained from the substitution of (\ref{eq:c+1}), (\ref{eq:c+2}),
(\ref{eq:c-1}), and (\ref{eq:c-2}) into (\ref{eq:co6}), (\ref{eq:c+6}), and (\ref{eq:c-6}):
\begin{align}
&(v_{1-})'''+2(v_{0-})''-w''_{1}v_{1-}-4(w_{1-})'(v_{1-})'+w_{1\cP}(v_{1-})''
 -2(w_{1+}+2w_{1-})'v_{0}\notag\\
&-2w_{1}(v_{0+})'+2(w_{0\cP})'v_{1-}+2(2w_{0+}-w_{0-})(v_{1-})'+4w_{0-}v_{0-}
 \notag\\
&-2\left[2w_{1-}(w_{1-})'+w_{1\cP}(w_{1+})'\right]v_{1-}-2w_{1}w_{1-}(v_{1-})'
 -4w_{1+}w_{1-}v_{0}\notag\\
&-4w_{1-}w_{0-}v_{1-}-2(v_{1-})^{2}(v_{1-})'-8C_{01}w_{0-}=0,
\label{eq:co6'}\\
&(v_{1-})'''+2(v_{0-})''+2(w_{1+})''v_{1-}-5(w_{1-})'(v_{1-})'-2w_{1+}(v_{1-})''
 -2(w_{1+})'v_{0}\notag\\
&-2(w_{1-})'(2v_{0+}-v_{0-})-2w_{1}(v_{0\cP})'+2(w_{0\cP})'v_{1-}
 +2(2w_{0+}-w_{0-})(v_{1-})'\notag\\
&+4w_{0-}v_{0-}-6w_{1-}(w_{1-})'v_{1-}-3\left[(w_{1+})^{2}+(w_{1-})^{2}\right]
 (v_{1-})'-2\left[ (w_{1+})^{2}+(w_{1-})^{2}\right] v_{0+}\notag\\
&+4w_{1-}w_{0+}v_{1-}-3(v_{1-})^{2}(v_{1-})'-2(v_{1-})^{2}v_{0+}-2\left[(w_{1+})^{2}
 +(w_{1-})^{2}\right] w_{1-}v_{1-}\notag\\
&-2w_{1-}(v_{1-})^{3}+8C_{11}=0,
\label{eq:c+6'}\\
&(v_{1-})'''+2(v_{0-})''-2(w_{1-})''v_{1-}-3(w_{1-})'(v_{1-})'-2w_{1-}(v_{1-})''
 -2w'_{1}v_{0}\notag\\
&-(2w_{1-})'v_{0+}-2w_{1}v'_{0}+2(w_{0\cP})'v_{1-}+2(2w_{0+}-w_{0-})(v_{1-})'
 +4w_{0-}v_{0-}\notag\\
&-2w_{1-}(w_{1-})'v_{1-}-\left[(w_{1+})^{2}+(w_{1-})^{2}\right] (v_{1-})'
 +2\left[(w_{1+})^{2}+(w_{1-})^{2}\right]v_{0+}\notag\\
&-4w_{1-}w_{0+}v_{1-}-(v_{1-})^{2}(v_{1-})'+2(v_{1-})^{2}v_{0+}+2\left[ (w_{1+})^{2}
 +(w_{1-})^{2}\right]w_{1-}v_{1-}\notag\\
&+2w_{1-}(v_{1-})^{3}-8C_{11}=0.
\label{eq:c-6'}
\end{align}
The even and odd parts of the condition (\ref{eq:co8}):
\begin{align}
&2w_{1+}(w_{1+})''+2w_{1-}(w_{1-})''-((w_{1+})')^{2}-((w_{1-})')^{2}+2v_{1-}(v_{1-})''
 \notag\\
&-((v_{1-})')^{2}+4(v_{0-})^{2}-2\left[(w_{1+})^{2}+(w_{1-})^{2}\right] (w_{1-})'
 -4w_{1+}w_{1-}(w_{1+})'\notag\\
&+4\left[(w_{1+})^{2}+(w_{1-})^{2}\right] w_{0+}+8w_{1+}w_{1-}w_{0-}
 -2(w_{1-})'(v_{1-})^{2}-4w_{1-}v_{1-}(v_{1-})'\notag\\
&-8w_{1-}v_{1-}v_{0+}+4w_{0+}(v_{1-})^{2}-(w_{1+})^{4}
 -6(w_{1+})^{2}(w_{1-})^{2}-(w_{1-})^{4}\notag\\
&-2\left[(w_{1+})^{2}+3(w_{1-})^{2}\right] (v_{1-})^{2}-(v_{1-})^{4}-16C_{10}=0,
\label{eq:co8+}\\
&(w_{1+})''w_{1-}-(w_{1+})'(w_{1-})'+w_{1+}(w_{1-})''-2w_{1+}w_{1-}(w_{1-})'
 -\left[(w_{1+})^{2}\right.\notag\\
&\left. +(w_{1-})^{2}\right] (w_{1+})'+2\left[(w_{1+})^{2}+(w_{1-})^{2}\right] 
 w_{0-}+4w_{1+}w_{1-}w_{0+}-w_{1+}v_{1-}(v_{1-})'\notag\\
&-2w_{1+}v_{1-}v_{0+}-2(w_{1+})^{3}w_{1-}-2w_{1+}(w_{1-})^{3}
 -2w_{1+}w_{1-}(v_{1-})^{2}=0,
\label{eq:co8-}
\end{align}
where (\ref{eq:c-4'}) and (\ref{eq:co7}) have been used to eliminate $(v_{0-})'$ in
(\ref{eq:co8-}).\\

\noindent
The even and odd parts of the condition (\ref{eq:co8'}):
\begin{align}
&(w_{1-})'''-w_{1+}(w_{1+})''-w_{1-}(w_{1-})''-2((w_{1+})')^{2}-2((w_{1-})')^{2}
+4(w_{1+})'w_{0-}\notag\\
&+2w_{1+}(w_{0-})'+4(w_{1-})'w_{0+}+2w_{1-}(w_{0+})'-v_{1-}(v_{1-})''
-2((v_{1-})')^{2}\notag\\
&-4(v_{1-})'v_{0+}-2v_{1-}(v_{0+})'-2\left[(w_{1+})^{2}+(w_{1-})^{2}\right]
 (w_{1-})'-4w_{1+}w_{1-}(w_{1+})'\notag\\
&-2(w_{1-})'(v_{1-})^{2}-4w_{1-}v_{1-}(v_{1-})'=0,
\label{eq:co8'+}\\
&(w_{1+})'''-(w_{1+})''w_{1-}-4(w_{1+})'(w_{1-})'-w_{1+}(w_{1-})''+4(w_{1+})'w_{0+}
 \notag\\
&+4(w_{1-})'w_{0-}+2w_{1+}(w_{0+})'+2w_{1-}(w_{0-})'+2(v_{1-})'v_{0-}+4v_{0+}v_{0-}
 \notag\\
&-2\left[(w_{1+})^{2}+(w_{1-})^{2}\right] (w_{1+})'
 -4w_{1+}w_{1-}(w_{1-})'-2(w_{1+})'(v_{1-})^{2}=0.
\label{eq:co8'-}
\end{align}
The even and odd parts of the condition (\ref{eq:co9}):
\begin{align}
&(w_{1-})''v_{1-}-(w_{1-})'(v_{1-})'+w_{1-}(v_{1-})''-\left[w_{1+}(w_{1+})'
 +2w_{1-}(w_{1-})'\right] v_{1-}\notag\\
&-(w_{1-})^{2}(v_{1-})'-2(w_{1-})^{2}
v_{0+}+2w_{1+}w_{0-}v_{1-}+4w_{1-}w_{0+}v_{1-}-(v_{1-})^{2}(v_{1-})'\notag\\
&-2(v_{1-})^{2}v_{0+}-2\left[(w_{1+})^{2}
+(w_{1-})^{2}\right] w_{1-}v_{1-}-2w_{1-}(v_{1-})^{3}+8C_{11}=0,
\label{eq:co9+}\\
&(w_{1+})''v_{1-}-w_{1+}(v_{1-})''+2(w_{1-})'v_{0-}+2w_{1-}(v_{0-})'=0,
\label{eq:co9-}
\end{align}
where (\ref{eq:c-4'}) and (\ref{eq:co7}) have been used to eliminate $(v_{0-})'$ in
(\ref{eq:co9+}).\\

\noindent
The even and odd parts of the condition (\ref{eq:co9'}):
\begin{align}
&(v_{1-})'''-(w_{1-})''v_{1-}-4(w_{1-})'(v_{1-})'-w_{1-}(v_{1-})''-2(w_{1+})'v_{0-}
 \notag\\
&-4(w_{1-})'v_{0+}-2w_{1-}(v_{0+})'+2(w_{0+})'v_{1-}+4w_{0+}(v_{1-})'+4w_{0-}v_{0-}
 \notag\\
&-4w_{1-}(w_{1-})'v_{1-}-2\left[(w_{1+})^{2}+(w_{1-})^{2}\right] (v_{1-})'
 -2(v_{1-})^{2}(v_{1-})'=0,
\label{eq:co9'+}\\
&2(v_{0-})''+(w_{1+})''v_{1-}-w_{1+}(v_{1-})''-2(w_{1+})'v_{0+}-2w_{1+}(v_{0+})'\notag\\
&-2(w_{0-})'v_{1-}-2w_{0-}(v_{1-})'=0.
\label{eq:co9'-}
\end{align}
The definition of $f_{i}$ ($i=1,2,3$):
\begin{align}
&f_{1}=-2w_{1+}(w_{1+})''-2w_{1-}(w_{1-})''+((w_{1+})')^{2}+((w_{1-})')^{2}
 -2v_{1-}(v_{1-})''\notag\\
&+((v_{1-})')^{2}-4(v_{0-})^{2}+2\left[(w_{1+})^{2}+(w_{1-})^{2}\right] (w_{1-})'
 +4w_{1+}w_{1-}(w_{1+})'\notag\\
&+2(w_{1-})'(v_{1-})^{2}+4w_{1-}v_{1-}(v_{1-})'+(w_{1+})^{4}+6(w_{1+})^{2}
 (w_{1-})^{2}+(w_{1-})^{4}\notag\\
&+2\left[(w_{1+})^{2}+3(w_{1-})^{2}\right] (v_{1-})^{2}+(v_{1-})^{4}+16C_{10},
\label{eq:f1}\\
&f_{2}=-(w_{1+})''w_{1-}+(w_{1+})'(w_{1-})'-w_{1+}(w_{1-})''+2w_{1+}w_{1-}(w_{1-})'
 \notag\\
&+\left[(w_{1+})^{2}+(w_{1-})^{2}\right](w_{1+})'+w_{1+}v_{1-}(v_{1-})'
 +2(w_{1+})^{3}w_{1-}\notag\\
&+2w_{1+}(w_{1-})^{3}+2w_{1+}w_{1-}(v_{1-})^{2},
\label{eq:f2}\\
&f_{3}=-(w_{1-})''v_{1-}+(w_{1-})'(v_{1-})'-w_{1-}(v_{1-})''+\left[w_{1+}(w_{1+})'
 \right.\notag\\
&\left. +2w_{1-}(w_{1-})'\right] v_{1-}+(w_{1-})^{2}(v_{1-})'+(v_{1-})^{2}(v_{1-})'
 +2\left[(w_{1+})^{2}\right.\notag\\
&\left. +(w_{1-})^{2}\right] w_{1-}v_{1-}+2w_{1-}(v_{1-})^{3}-8C_{11}.
\label{eq:f3}
\end{align}

%\section*{References}%IOPART

\bibliography{refsels}%BIB-FILE
\bibliographystyle{npb}%BST-FILE
%\begin{thebibliography}{99}
%\def\J#1#2#3#4{{\sl #1} {\bf #2} (#3) #4}

%\bibitem{}
%Author 1, Author 2 and Author 3,
%\J{Journal}{Volume}{Year}{Page}.

%\end{thebibliography}

\end{document}